\documentclass[aps,pre,twocolumn,showpacs]{revtex4}

\usepackage{amssymb,amsmath,graphicx}

\begin{document}

\title{Variable-delay feedback control of unstable steady
states \\ in retarded time-delayed systems}

\author{A. Gjurchinovski}

\email{agjurcin@pmf.ukim.mk}

\author{V. Urumov}

\affiliation{Institute of Physics, Faculty of Natural Sciences
and Mathematics, Sts.\ Cyril and Methodius University,
P.\ O.\ Box 162, 1000 Skopje, Macedonia}


\begin{abstract}
We study the stability of unstable steady states in scalar 
retarded time-delayed systems subjected to a variable-delay 
feedback control. 
The important aspect of such a control problem is that 
time-delayed systems are already infinite-dimensional before 
the delayed feedback control is turned on. 
When the frequency of the modulation is large compared to the 
system's dynamics, the analytic approach consists of relating 
the stability properties of the resulting variable-delay system 
with those of an analogous distributed delay system. 
Otherwise, the stability domains are obtained by a numerical 
integration of the linearized variable-delay system.
The analysis shows that the control domains are significantly 
larger than those in the usual time-delayed feedback control, 
and that the complexity of the domain structure depends on 
the form and the frequency of the delay modulation. 
\end{abstract}

\date{24 December, 2009}

\maketitle

\section{Introduction}

In spite of the fact that the control problems have been
thoroughly investigated from a theoretical aspect 
and the results are being implemented in 
concrete real systems for several decades \cite{OGA97}, the control 
of chaotic dynamical systems is a relatively new area 
of research. The appearance of the pioneering 
paper by Ott, Grebogy and Yorke (OGY) in 1990 
boosted quite an interest among nonlinear scientists \cite{OGY90}, 
being a reason for a large number of published papers on 
chaos control \cite{SCH08,BGL00,KAP96,ABC98}. 
The OGY method utilizes the existence of 
infinitely many unstable periodic orbits (UPO) within 
the structure of the chaotic attractor, applying a small 
externally controlled perturbation to suitably chosen 
parameters of the system when the trajectory 
is in the neighborhood of an UPO whose control 
is desirable. The system is then externally forced to 
follow otherwise unstable behavior corresponding to that 
UPO. The numerical simulations and the experimental 
implementations showed that the method by 
itself has some drawbacks concerning the robustness 
with respect to the external noise and its practical 
realization, since it requires a continuous monitoring 
of the evolution of the system from the outside and the 
knowledge of the equations that describe the system's dynamics.    

The OGY idea stimulated a development of a rich 
variety of new chaos control techniques. 
Among those is the time-delayed feedback control (TDFC) 
proposed by Pyragas in 1992 \cite{PYR92,PYR06}, shown to be much more 
flexible for practical purposes with respect to OGY 
(the monitoring of the system and the knowledge of 
the exact positions of UPOs are not required) and 
quite robust against the effects of noise. The control 
force is applied as a continuous feedback proportional to 
the difference between the current state of the system and the 
state of the system delayed by the constant time $T$. 
If the time delay $T$ is chosen to coincide with an 
integer multiple of the period of the target UPO, then the 
control force will vanish when the target state is reached 
and the control is noninvasive. For stabilization of 
unstable steady states (USS), the choice of the parameter 
$T$ is not as restrictive as in the case of UPOs, and the 
interval of $T$ for which TDFC is successful is shown to be 
system-dependent \cite{HS05,YWH06,DHS07}. 

In parallel to various practical applications of TDFC
\cite{PT93,BDG94,PBA96,HCT97,SBG97,KF00,FSK02,RP04,LBJ04,PHT05}, 
an effort has been put into progress to generalize or 
modify the original control scheme in order to improve 
its performance. Some extended TDFC schemes employ multiple 
time-delays to stabilize strongly unstable periodic orbits
\cite{SSG94,PYR95,AP04,AP05}. Others are introducing 
an oscillating feedback gain \cite{SS97} or 
an extra unstable degree of freedom in the feedback loop
\cite{P01,PPK02,PPK04} to overcome the so-called odd-number 
limitation \cite{JBO97,NAK97,NU98}, which was refuted 
recently \cite{FFG07,JFG07,PS07,FYF08}. In a recent work \cite{GU08}, 
it has been shown that the efficiency of the TDFC 
method to control USS can be significantly improved by 
including a variable time-delay into the TDFC scheme in 
a form of a deterministic or stochastic modulation in a 
fixed interval around a nominal delay value. 
Stochastic changes in the delay time are natural due to 
the omnipresent noise in any physical system. In the 
circumstances, the enhancement of noise along the delay 
line could be desirable as it is leading to improved 
stability of the system. On the other hand the modulated 
delay described by some periodic function could be realized 
by periodically changing some characteristic distances 
in electric or laser systems by introducing piezoelements.  
This variable delay feedback control (VDFC) has been 
shown successful in stabilization of USS in 
low-dimensional chaotic systems using different types 
of delay modulations. The ongoing analysis shows that 
VDFC can also improve the control domain of UPOs 
with respect to TDFC for a specific choice of the delay 
modulation \cite{GU09}. 

The purpose of this paper is to investigate the 
effects of stabilization of unstable equilibria 
by a variable-delay feedback control in a class of 
nonlinear dynamical systems described by scalar 
retarded delay-differential equations (RDDE) 
involving the value of the state variable at 
a discrete time lag. A delay differential 
equation is called retarded if the highest order 
derivative only occurs with one value of the argument, 
and this argument is not less than the arguments of the 
unknown function and its lower order derivatives 
appearing in the equation \cite{BC63,HAL71,BZ03}. In contrast to low 
dimensional dynamical systems, delay differential 
equations are infinite dimensional, since it is 
necessary to specify a continuum of initial 
conditions over the interval length equal to the 
time delay. 
The interest for such equations is caused by 
their frequent occurrence in numerous physical, biological 
and engineering models, where the time delays are a natural
manifestation of the system's dynamics \cite{WY06,BKH08,ERN09,YP09}.

The paper is organized as follows. In Section II 
we perform a linear stability analysis of USS in the 
free running RDDE system and in the system 
under VDFC. The frequency of the delay modulation in 
the feedback loop is considered to be 
sufficiently large compared to the intrinsic timescale 
of the unperturbed system, allowing an approximation 
of the variable delay system with a distributed 
delay system \cite{MAN05}. 
In Section III, we numerically illustrate the VDFC 
method in the chaotic Mackey-Glass system. 
The domains of successful control are first computed for 
high-frequency modulations of the time delay for different values 
of the modulation amplitude.
The planes of the control domains are parametrized by the 
feedback gain and the nominal delay of 
the control force for a fixed delay of the RDDE, and also,
by the time delay of the original 
system and the nominal delay of the feedback control force
for a fixed value of the feedback gain.
The control domains are also determined for a 
low-frequency modulation in the plane of the feedback gain 
and the nominal delay of the control force for different 
values of the frequency of the modulation. 
The results show a significant enlargement of stability
areas of VDFC with respect to TDFC
within a certain range of the control parameters, 
sometimes resulting in a complicated reconfiguration 
depending on the type, the amplitude and the frequency 
of the delay modulation.
The conclusions are summarized in Section IV.

\section{Stability analysis}

We consider a general nonlinear dynamical 
system described by a scalar autonomous RDDE in the form:
\begin{equation}
\dot{x}(t)=F[x(t),x(t-T_1)],
\label{2.1}
\end{equation}
where $T_1\geq0$ is a constant delay time, and $F$ is an arbitrary 
nonlinear function of the state variable $x$, having a past 
dependence through the same state variable $x$ but at $T_1$ 
time units in the past. The presence of the delay term $x(t-T_1)$ 
is a cause for the system (\ref{2.1}) to be infinite dimensional, 
since a continuum of initial conditions over the time interval 
$[-T_1,0]$ is required in order to uniquely specify the future behavior 
of the system. The system possesses a set of fixed points $\{x^*_i\}$ 
that are solution of $F[x^*(t),x^*(t-T_1)]=0$, and the stability 
of a particular fixed point $x^*$ can be obtained
by linearizing Eq. (\ref{2.1}) in the vicinity of $x^*$.
The linearized version of (\ref{2.1}) around $x^*$ has a 
general form:
\begin{equation}
\dot{\widetilde{x}}(t)=A\,\widetilde{x}(t)+B\,\widetilde{x}(t-T_1),
\label{2.2}
\end{equation}
where $A$ and $B$ are real constants. We made a coordinate 
transformation from $x$ to $\widetilde{x}$ according to 
$\widetilde{x}(t)=x(t)-x^*$ such that the fixed point is at 
the origin as expressed in the new coordinate. Employing 
the usual ansatz $x(t)\sim\exp(\lambda t)$ in (\ref{2.2}) 
we obtain the characteristic equation:
\begin{equation}
\lambda=A+B\,e^{-\lambda T_1}.
\label{2.3}
\end{equation}
This is a transcendental equation in $\lambda$, possessing
a countable infinite set of complex solutions $\{\lambda_i\}$
defining the eigenvalues of the fixed point at the origin.
The origin is stable if and only if each $\lambda_i$ has a 
negative real part, it is unstable if at least one $\lambda_i$ 
has a positive real part, and it is marginally unstable 
if the largest real part of all the eigenvalues $\{\lambda_i\}$ 
is zero. 

The goal of this paper is to investigate the possibility of 
stabilization of the unstable fixed point $x^*$ of the 
system (\ref{2.1}) by applying a Pyragas-type feedback 
force $u(t)$ with a variable time delay \cite{GU08}:
\begin{eqnarray}
u(t)&=&K\,[x(t-\tau(t))-x(t)], \label{2.4}\\
\tau(t)&=&T_2+\varepsilon\,f(\nu t),
\label{2.5}
\end{eqnarray}
such that for a given set of control parameters $\{K,T_2,\varepsilon,\nu\}$
the unstable fixed point $x^*$ of the unperturbed system 
(\ref{2.1}) becomes stable in the presence of the feedback 
term (\ref{2.4}). The control parameter $K$ is the feedback 
gain characterizing the strength of the feedback, and $\tau(t)$ 
is the variable time delay. We will consider a variation in a form of a 
deterministic modulation around a nominal delay value described by the 
control parameter $T_2$. We take the delay function 
$f:\mathbb{R}\rightarrow[-1,1]$ to be periodic with 
zero mean, with $\varepsilon$ and $\nu$ being the parameters 
determining the amplitude and the frequency of the modulation, 
respectively. 
The form of the control force (\ref{2.4})--(\ref{2.5}) implies that
since $\tau(t)\geq0$, the values of the amplitude $\varepsilon$
are restricted to the interval $[0,T_2]$.
In the presence of the control force (\ref{2.4}), 
the system (\ref{2.1}) has the form:
\begin{equation}
\dot{x}(t)=F[x(t),x(t-T_1)]+u(t),
\label{2.6}
\end{equation}
and the linearized version around $x^*$ in terms of the 
new coordinate $\widetilde{x}$ is:
\begin{equation}
\dot{\widetilde{x}}(t)=A\,\widetilde{x}(t)+B\,\widetilde{x}(t-T_1)+\widetilde{u}(t),
\label{2.7}
\end{equation}
where
\begin{equation}
\widetilde{u}(t)=K\,[\widetilde{x}(t-\tau(t))-\widetilde{x}(t)].
\label{2.8}
\end{equation}

The stability of the origin can be inferred by numerically integrating
the linear variable-delay system (\ref{2.7})--(\ref{2.8}) for different
values of $K$, $T_2$, $\varepsilon$ and $\nu$, thus determining the domains in the
$(K,T_2,\varepsilon,\nu)$ hyperspace for which the stabilization becomes 
possible. 

For a sufficiently large variation of the time delay $\tau(t)$, 
the stability of the linear variable-delay system (\ref{2.7})--(\ref{2.8}) 
becomes amenable for analytical treatment  \cite{MAN05}. From the stability 
point of view, if the frequency of the delay variation $\nu$ is sufficiently 
large, then the linear system (\ref{2.7}) with a variable time-delay 
(\ref{2.8}) behaves as the following time-invariant system with a 
distributed delay (Theorem A1, Appendix A):
\begin{eqnarray}
\dot{\widetilde{x}}(t)&=&A\,\widetilde{x}(t)+B\,\widetilde{x}(t-T_1)+ \nonumber\\
& &K\,\left(\int_{-1}^{1}w(\eta)\,\widetilde{x}(\varepsilon\eta+t-T_2)\,d\eta-\widetilde{x}(t)\right),
\label{2.9}
\end{eqnarray}
with $w$ being the weight related to the probability distribution of 
the delay function $f$ in the interval of its periodicity, satisfying $\int_{-1}^{1} 
w(\eta)\,d\eta=1$ (see Table I).
\begin{table*}
\caption{A representation of the delay function $f$, the weight $w$ 
of the distributed delay system, and the function $g$, corresponding 
to three different types of delay modulations.
By $I_0$ we denote the modified Bessel function of the first kind of order zero, 
$J_0$ is the Bessel function of the first kind of order zero, and $\delta$ is 
the Dirac delta function.}

\begin{tabular}{l|c|c|c|c}

\hline\hline
\multicolumn{1}{c|}{Type} & 
\multicolumn{1}{c|}{$f(t)$} & 
\multicolumn{1}{c|}{$w(t)$} & 
\multicolumn{1}{c|}{$g(\lambda\varepsilon)$} & 
\multicolumn{1}{|c}{$g(i\omega\varepsilon)$}  \\

\hline\hline
&&&&\\
Sawtooth wave & 
$
\left\{
\begin{array}{cc}
\displaystyle{\frac{2}{\pi}\left(t-\frac{\pi}{2}\right),} & t\in[0,\pi) \\
&\\
\displaystyle{\frac{2}{\pi}\left(\frac{3\pi}{2}-t\right),} & t\in[\pi,2\pi) \\
\end{array}\right.
$ &
$\displaystyle{\frac{1}{2}}$ &
$\displaystyle{\frac{\sinh(\lambda\varepsilon)}{\lambda\varepsilon}}$ &
$\displaystyle{\frac{\sin(\omega\varepsilon)}{\omega\varepsilon}}$ \\
&&&&\\
\hline

&&&&\\
Sine wave & 
$
\sin(t)
$ &
$\displaystyle{\frac{1}{\pi\sqrt{1-t^2}}}$ &
$I_0(\lambda\varepsilon)$ &
$J_0(\omega\varepsilon)$ \\
&&&&\\

\hline
&&&&\\
Square wave & 
$
\left\{
\begin{array}{cc}
-1, & t\in[0,\pi) \\
1, & t\in[\pi,2\pi) \\
\end{array}\right.
$ &
$\displaystyle{\frac{\delta(t-1)+\delta(t+1)}{2}}$ &
$\cosh(\lambda\varepsilon)$ &
$\cos(\omega\varepsilon)$ \\
&&&&\\
\hline\hline
\end{tabular}

\end{table*}
The stability of the distributed delay system (\ref{2.9}) is 
determined by the roots $\lambda_i$ of its characteristic 
equation:
\begin{equation}
\lambda=A+B\,e^{-\lambda T_1}+K\,\left[e^{-\lambda T_2}\,g(\lambda\varepsilon)-1\right],
\label{2.10}
\end{equation}
where $g:\mathbb{C}\rightarrow\mathbb{C}$ is a smooth complex function defined as:
\begin{equation}
g(\lambda\epsilon)=\int_{-1}^{1}w(\eta)e^{\lambda\epsilon\eta}d\eta.
\label{eq:2.11}
\end{equation}
In this sense, the solutions $\{\lambda_i\}$ determining the stability of the
comparison system (\ref{2.9}) can be considered as effective 
eigenvalues describing the overall stability of the original 
variable delay system (\ref{2.7})--(\ref{2.8}), providing that the 
delay frequency $\nu$ is large compared to the system's dynamics.
Numerical simulations showed that the threshold for the frequency $\nu$ 
above which this type of comparative analysis becomes valid needs not to be
very high, and that its value depends on the actual system under 
investigation.

\subsection{Stability of the unperturbed system}

In the absence of control, the stability of the fixed point $x^*$ 
is determined by the roots of the characteristic equation (\ref{2.3}).
Let $H_0$ be a function of $\lambda$ defined as:
\begin{equation}
H_0(\lambda)=\lambda-A-B\,e^{-\lambda T_1}.
\label{2.12}
\end{equation}
With the aid of this characteristic quasipolinomial $H_0(\lambda)$, 
Eq. (\ref{2.3}) can be written as $H_0(\lambda)=0$. 
We would like to find the range of the values for $A$, $B$ and $T_1$ 
for which $x^*$ is stable. 

Since $H_0(\lambda)$ is a smooth function on $\lambda$, it is useful to
consider the behavior of $H_0(\lambda)$ as $\lambda$ changes continuously 
over the real interval $[0,+\infty)$. Specifically, at the ends
of this interval, we have:
\begin{eqnarray}
\lim_{\lambda\rightarrow\infty}H_0(\lambda)&=&+\infty,
\label{2.13}\\
\lim_{\lambda\rightarrow 0^+}H_0(\lambda)&=&-(A+B).
\label{2.14}
\end{eqnarray}
If $A+B>0$, then $H_0$ changes its sign at least once as $\lambda$
sweeps along the positive real axis. Consequently, there exists 
at least one positive real root of the characteristic equation
$H_0(\lambda)=0$, rendering the fixed point unstable 
for any $T_1$. If $A+B=0$, then $\lambda=0$ is a root of the
characteristic equation (\ref{2.3}), and the fixed point is unstable,
or at least marginally unstable. Hence, a necessary (but not sufficient!)
condition for stability of the fixed point is:
\begin{equation}
A+B<0.
\label{2.15}
\end{equation}

Taking into account that the boundary between stability
and instability (the threshold of control) occurs when 
the maximal value from all the real parts in the set of solutions
$\{\lambda_i\}$ is zero, we look for a solution of Eq. (\ref{2.3}) 
in the form $\lambda=i\omega$, $\omega\in\mathbb{R}$,
and separate real and imaginary parts of the resulting equation
to obtain:
\begin{eqnarray}
-A&=&B\cos(\omega T_1), 
\label{2.16}\\
-\omega&=&B\sin(\omega T_1).
\label{2.17}
\end{eqnarray}
[We stress that a zero on the imaginary axis for some set of parameters 
$A$, $B$ and $T_1$ does not necessarily mean that all the other
zeros of the characteristic polinomial $H_0(\lambda)$ for 
the same set of parameters have negative real parts. 
The stability boundary is just one set of solutions of Eqs. 
(\ref{2.16})--(\ref{2.17}).]
By eliminating the trigonometric terms from the last pair of
equations, we get:
\begin{equation}
\omega^2=B^2-A^2,
\label{2.18}
\end{equation}
from which we conclude that Eq. (\ref{2.3}) can have a solution 
for $\lambda$ on the imaginary axis if and only if $|B|>|A|$.
Taking into account that $T_1>0$, from Eq. (\ref{2.16}) we obtain:
\begin{equation}
T_1=\frac{\mathrm{Arccos}(-A/B)+2n\pi}{\sqrt{B^2-A^2}},
\label{2.19}
\end{equation}
where $n$ is a nonnegative integer, and Arccos denotes the principal value
of the arccosine function. Obviously, the first value of $T_1$
for which $H_0(\lambda)=0$ has a solution for $\lambda$ on the imaginary
axis is:
\begin{equation}
T_1^*=\frac{\mathrm{Arccos}(-A/B)}{\sqrt{B^2-A^2}},
\label{2.20}
\end{equation}
which follows from Eq. (\ref{2.19}) by setting $n=0$. The behavior of the
real part of $\lambda$ at the values for $T_1$ in Eq. (\ref{2.19})
is determined by the derivative $d\lambda/dT_1$ at $\lambda=i\omega$.
By implicit differentiation of Eq. (\ref{2.3}) with respect to
$T_1$, we obtain:
\begin{equation}
\frac{d\lambda}{dT_1}=-\frac{\lambda B\,e^{-\lambda T_1}}{1+BT_1\,e^{-\lambda T_1}}
=-\frac{\lambda(\lambda-A)}{1+T_1(\lambda-A)},
\label{2.21}
\end{equation}
from which at $\lambda=i\omega$ we get:
\begin{equation}
\mathrm{Re}\,\left(\frac{d\lambda}{dT_1}\right)_{\lambda=i\omega}=
\frac{\omega^2}{(1-AT_1)^2+(\omega T_1)^2}.
\label{2.22}
\end{equation}
Since the sign of this derivative is always positive, 
the sign of the real part of $\lambda$ switches from negative to
positive when the zero of the characteristic quasipolinomial 
$H_0(\lambda)$ crosses the imaginary axis. On the other hand, as an 
implication of the Rouch\'{e} theorem, the number of roots 
(counting multiplicity) on the complex right half plane (RHP) 
and the number of roots on the complex left half plane (LHP) 
can be changed (or, more correctly, interchanged) only if a 
zero appears on or crosses the imaginary axis \cite{HEN74,GNC05}. 
As a consequence, 
in the case under consideration $|B|>|A|$, all the zeros of 
the characteristic quasipolinomial $H_0(\lambda)$ lie on the 
LHP if $T_1$ is in the interval $[0,T_1^*)$ providing that 
all the zeros were on the LHP before the first crossing of 
the imaginary axis has occured. However, this is evidently 
not true for other intervals separated by the corresponding 
values of $T_1$ given by Eq. (\ref{2.19}) for $n>0$, since 
the first zero-crossing of the imaginary axis occurs for 
$T_1=T_1^*$, and according to Eq. (\ref{2.22}) every crossing is 
from the LHP to the RHP.

In the case $A\geq0$, the necessary condition for 
the stability of the fixed point $x^*$ is $B<-A$ [see Eq. (\ref{2.15})], 
which is an interval of $B$ that belongs to the range 
$|B|>|A|$ for which the characteristic equation 
(\ref{2.3}) can have a solution on the imaginary axis. 
From the previous discussion, the possibility for all 
the zeros of the quasipolinomial $H_0(\lambda)$ to lie on the LHP 
necessary imply $T_1\in[0,T_1^*)$. 
Since for $T_1=0$ the characteristic equation (\ref{2.3}) 
is reduced to $\lambda=A+B<0$, and since the crossing of the 
imaginary axis occurs for $T_1=T_1^*$, we conclude that all the
zeros $\{\lambda_i\}$ have negative real parts in this case
if and only if $B<-A$ and $T_1\in[0,T_1^*)$.

In the case $A<0$, the necessary condition for the stability
of the fixed point is $B<|A|$. In the subinterval 
$B\in[-|A|,|A|)$, the characteristic quasipolinomial 
(\ref{2.12}) cannot have a zero on the imaginary axis. 
Choosing $B=0$, from (\ref{2.3}) we have $\lambda=A<0$.
Since crossing of the imaginary axis does not occur
for this subinterval of $B$, it follows that all the zeros 
$\{\lambda_i\}$ for $B\in[-|A|,|A|)$ lie on the LHP for any
$T_1>0$. On the other hand, in the range $B<-|A|$ the 
characteristic quasipolinomial (\ref{2.12}) can 
have a zero on the imaginary axis. Putting $T_1=0$ 
in (\ref{2.3}) we obtain $\lambda=A+B<0$, which means 
that when $B<-|A|$ all the zeros $\{\lambda_i\}$ have 
negative real parts when $T_1\in[0,T_1^*)$.

The results are summarized with the following theorem:
\newtheorem{theorem}{Theorem}
\begin{theorem}
Let the linear RDDE:
\begin{equation}
\dot{\widetilde{x}}(t)=A\,\widetilde{x}(t)+B\,\widetilde{x}(t-T_1),
\hspace{1cm} A,B\in\mathbb{R}
\nonumber
\end{equation}
be a result of linearization of a corresponding nonlinear 
RDDE with a constant delay $T_1$:
\begin{equation}
\dot{x}(t)=F[x(t),x(t-T_1)]
\nonumber
\end{equation}
around some fixed point $x^*$ of the latter expressed in
coordinates in which the fixed point is at the origin.
Furthermore, let $T_1^*>0$ be a real positive constant defined as:
\begin{equation}
T_1^*=\frac{\mathrm{Arccos}(-A/B)}{\sqrt{B^2-A^2}}.
\nonumber
\end{equation}
Then, the fixed point $x^*$ is locally asymptotically stable in 
each of the following cases:\\

(a). $B<-|A|$ and $T_1\in[0,T_1^*)$;
\smallskip

(b). $B\in[-|A|,|A|)$, $A<0$ and $T_1>0$.\\

\noindent Otherwise, $x^*$ is unstable.
\end{theorem} 

\subsection{Stability under variable-delay feedback control 
(high-frequency modulation)}

In the following, we consider the modulation frequency $\nu$ 
to be above the threshold, allowing an analysis of the 
variable delay system (\ref{2.7})--(\ref{2.8}) 
as a distributed delay system (\ref{2.9}).
When the control is switched on, 
the stability of the fixed point $x^*$ is determined by 
the roots $\{\lambda_i\}$ of the characteristic equation 
(\ref{2.10}). If we define:
\begin{equation}
H_\varepsilon(\lambda)=\lambda-A-B\,e^{-\lambda T_1}+
K\,\left[1-e^{-\lambda T_2}g(\lambda\varepsilon)\right],
\label{2.23}
\end{equation}
then Eq. (\ref{2.10}) can be rewritten as $H_\varepsilon(\lambda)=0$.
Assuming that in the absence of control, the parameters $A$, $B$ and
$T_1$ of the unperturbed system are such that $x^*$ is unstable,
we look for the values of the control parameters $K$,
$T_2$ and $\varepsilon$ for which the fixed point is stabilized.
In other words, we would like to find the set of points 
(i. e. to determine the domain of control) in the 
parameter space $(K,T_2,\varepsilon)$ for 
which all the zeros of the characteristic quasipolynomial 
$H_\varepsilon(\lambda)$
lie on the LHP, while, at the same time, the characteristic
quasipolynomial $H_0(\lambda)$ of the unperturbed system has at
least one zero in the RHP. 

Before we proceed with the analytical description of the control boundaries,
it is interesting to consider the behavior of $H_\varepsilon(\lambda)$
as $\lambda$ changes continuously over the positive real axis.
Taking into account that $g(0)=\int_{-1}^1 w(\eta)\,d\eta=1$,
from Eq. (\ref{2.23}) we obtain:
\begin{eqnarray}
\lim_{\lambda\rightarrow\infty}H_\varepsilon(\lambda)&=&+\infty,
\label{2.24}\\
\lim_{\lambda\rightarrow 0^+}H_\varepsilon(\lambda)&=&-(A+B),
\label{2.25}
\end{eqnarray}
which coincide with the limits (\ref{2.13})--(\ref{2.14}) for the
characteristic polynomial $H_0(\lambda)$ of the
unperturbed system, leading to the same necessary condition (\ref{2.15}) 
for stability of the fixed point. Since (\ref{2.15}) does not 
include the dependence on the control parameters
$K$, $T_2$ and $\varepsilon$, we conclude that VDFC is unsuccessful
for any values of the control parameters if the linearized version (\ref{2.2}) 
of the unperturbed system around $x^*$ is such that $A+B>0$.
This important result is expressed in the following theorem.
\begin{theorem}
Let
$\dot{\widetilde{x}}(t)=A\,\widetilde{x}(t)+B\,\widetilde{x}(t-T_1),
\,A,B\in\mathbb{R},$
be a linearization around the fixed point $x^*$ of 
the corresponding nonlinear RDDE with a constant delay $T_1$.
If $A+B>0$, then the variable-delay feedback control (\ref{2.4})--(\ref{2.5})
cannot stabilize the unstable fixed point $x^*$ for any 
value of the control parameters $K$, $T_2$ and $\varepsilon$.
\end{theorem} 
The limitation of the VDFC method imposed by Theorem 2 is 
a kind of an analogue to the odd-number limitation \cite{JBO97,NAK97,NU98}
in the case of delayed feedback control of systems described by ordinary 
differential equations, whose validity was recently refuted
\cite{FFG07,JFG07,PS07,FYF08} for the case of unstable periodic orbits.

Exact analytical description of the domains of successful control
in the parameter space $(K,T_2,\varepsilon)$ is difficult for the
characteristic Eq. (\ref{2.10}) due to the complexity
of the terms involving the dependence on $\lambda$. Thus, one should 
solve Eq. (\ref{2.10}) numerically in order to calculate the 
control domains. To this extend, it is possible to obtain 
expressions for the parametric representation of the control boundaries 
parametrized by a Hopf frequency $\omega$. Substituting $\lambda=i\omega$
in Eq. (\ref{2.10}) and separating real and imaginary parts, we
obtain:
\begin{eqnarray}
Kg(i\omega\varepsilon)\cos\omega T_2&=&K-A-B\cos\omega T_1, 
\label{2.26}\\
Kg(i\omega\varepsilon)\sin\omega T_2&=&-\omega-B\sin\omega T_1.
\label{2.27}
\end{eqnarray}
Elimination of $T_2$ from the last pair of equation yields
a quadratic equation in $K$:
\begin{eqnarray}
\lefteqn{\left[1-g(i\omega\varepsilon)^2\right]\,K^2-2(A+B\cos\omega T_1)\,K}\nonumber\\
& &+(A+B\cos\omega T_1)^2+(\omega+B\sin\omega T_1)^2=0,
\label{2.28}
\end{eqnarray}
which can be solved for $K$ in terms of $\omega$ to get:
\begin{eqnarray}
K(\omega)&=&\lefteqn{\frac{A+B\cos\omega T_1}{1-\left[g(i\omega\varepsilon)\right]^2}\pm
\frac{1}{1-\left[g(i\omega\varepsilon)\right]^2}}\nonumber\\
&\times&\left[\left[g(i\omega\varepsilon)\right]^2(A+B\cos\omega T_1)^2\right.\nonumber\\
&+&\left.(\left[g(i\omega\varepsilon)\right]^2-1)(\omega+B\sin\omega T_1)^2\right]^{1/2}.
\label{2.29}
\end{eqnarray}
On the other hand, by dividing (\ref{2.27}) and (\ref{2.26}), we obtain:
\begin{equation}
T_2(\omega)=\frac{1}{\omega}\left[\mathrm{Arctan}\left(\frac{-\omega-B\sin\omega T_1}{K-A-B\cos\omega T_1}\right)
\pm n\pi\right],
\label{2.30}
\end{equation}
which, together with Eq. (\ref{2.29}), describe the stability boundary 
for a fixed $\varepsilon$ in the $(K,T_2)$ plane, parametrized by $\omega$.

It is also useful to study the stability boundaries of the 
controlled system for a fixed feedback strength 
$K$ and modulation amplitude $\varepsilon$ in the 
parameter plane of the two delay times $(T_1,T_2)$. 
Following the idea in Ref. \cite{GNC05}, we rewrite 
Eq. (\ref{2.10}) as:
\begin{equation}
1+a(\lambda)e^{-\lambda T_1}+b(\lambda)e^{-\lambda T_2}=0,
\label{2.31}
\end{equation}
where $a(\lambda)$ and $b(\lambda)$ are given by:
\begin{eqnarray}
a(\lambda)&=&\frac{B}{A-K-\lambda},
\label{2.32}\\
b(\lambda)&=&\frac{K\,g(\lambda\varepsilon)}{A-K-\lambda}.
\label{2.33}
\end{eqnarray}
At the control boundary ($\lambda=i\omega$) the three terms in Eq. (\ref{2.31})
can be considered as three vectors in the complex plane, with the 
corresponding magnitudes 1, $|a(i\omega)|$ and $|b(i\omega)|$.
According to Eq. (\ref{2.31}), the sum of these vectors 
is a zero vector, thus forming the triangle shown in Fig. 1.
\begin{figure}
\includegraphics[width=0.9\columnwidth,height=!]{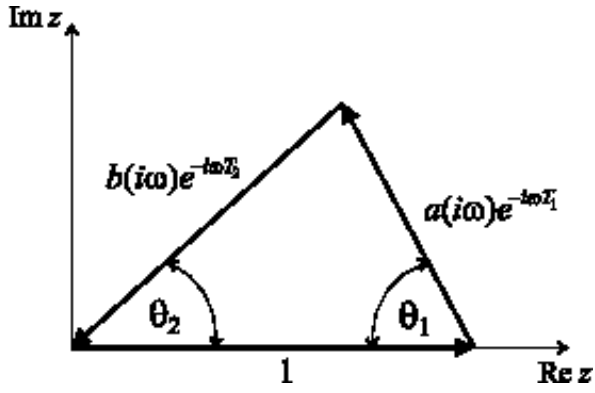}
\caption{Diagram in the complex plane $z$, associated
with the derivation of the parametric representation of the
stability boundary in ($T_1,T_2$) plane.}
\end{figure}
From Fig. 1, it is straightforward to obtain the parametric representation
of $T_1$ and $T_2$ on the Hopf frequency $\omega$:
\begin{eqnarray}
T_1(\omega)&=&\frac{\mathrm{Arg}\left[a(i\omega)\right]+(2u-1)\pi\pm\theta_1}{\omega}\geq 0,\nonumber\\
&&u=u_0^\pm,u_0^\pm+1,u_0^\pm+2\dots,\label{2.34}\\ \nonumber\\
T_2(\omega)&=&\frac{\mathrm{Arg}\left[b(i\omega\right]+(2v-1)\pi\mp\theta_2}{\omega}\geq 0,\nonumber\\
&&v=v_0^\pm,v_0^\pm+1,v_0^\pm+2\dots,\label{2.35},
\end{eqnarray}
where $u_0^\pm$ and $v_0^\pm$ are the smallest possible integers
such that the corresponding values of $T_1$ and $T_2$ are all non-negative, 
and $\theta_1,\theta_2\in[0,\pi]$ are the internal angles of the 
triangle shown in Fig. 1 calculated from the law of cosines as:
\begin{eqnarray}
\theta_1=\mathrm{Arccos}\left(\frac{1+|a(i\omega)|^2-|b(i\omega)|^2}{2|a(i\omega)|}\right),
\label{2.36}\\
\theta_2=\mathrm{Arccos}\left(\frac{1+|b(i\omega)|^2-|a(i\omega)|^2}{2|b(i\omega)|}\right).
\label{2.37}
\end{eqnarray}  

In the case when the nominal delay $T_2$ of the feedback control force
coincides with the delay of the original system $T_1$, the characteristic
Eq. (\ref{2.10}) is reduced to:
\begin{equation}
\lambda-A+K-\left[B+Kg(\lambda\varepsilon)\right]\,e^{-\lambda T}=0,
\label{2.38}
\end{equation}
where we use $T=T_1=T_2$. At the stability boundary ($\lambda=i\omega$)
the last complex equation can be represented as a pair of two 
real equations:
\begin{eqnarray}
\left[B+Kg(i\omega\varepsilon)\right]\cos\omega T&=&K-A,
\label{2.39}\\
\left[B+Kg(i\omega\varepsilon)\right]\sin\omega T&=&-\omega,
\label{2.40}
\end{eqnarray}
which can be manipulated to obtain a parametric representation
of the control boundary in the $(K,T)$ plane in terms of $\omega$:
\begin{eqnarray}
K(\omega)&=&\frac{A+Bg(i\omega\varepsilon)}{1-\left[g(i\omega\varepsilon)\right]^2}\pm
\frac{1}{1-\left[g(i\omega\varepsilon)\right]^2}\left[\left(A+Bg(i\omega\varepsilon)\right)^2\right.\lefteqn{}\nonumber\\
& &+\left.(\left[g(i\omega\varepsilon)\right]^2-1)(A^2+B^2-\omega^2)\right]^{1/2},
\label{2.41}\\
T(\omega)&=&\frac{1}{\omega}\left[\mathrm{Arctan}\left(\frac{-\omega}{K-A}\right)
\pm n\pi\right].
\label{2.42}
\end{eqnarray}

When $\varepsilon=0$, VDFC reduces to the usual Pyragas
control scheme (TDFC) with a constant delay $T_2$. Since 
TDFC is a special case of VDFC when the modulation of the 
control delay in the feedback force is absent, the 
parametric representations of the control boundaries 
for TDFC simply follow from the 
ones derived in the case of VDFC by letting 
$\varepsilon=0$ (or, equivalently, $g(0)=1$) in the corresponding equations.
For example, from Eqs. (\ref{2.26})--(\ref{2.27}) with $\varepsilon=0$
we obtain the parametric representation of the TDFC boundary
in the $(K,T_2)$ plane parametrized by $\omega$:
\begin{eqnarray}
K(\omega)=\frac{(A+B\cos\omega T_1)^2+(\omega+B\sin\omega T_1)^2}{2(A+B\cos\omega T_1)},\lefteqn{}
\label{2.43}\\
T_2(\omega)=\frac{1}{\omega}\left[\mathrm{Arctan}\left(\frac{-\omega-B\sin\omega T_1}{K-A-B\cos\omega T_1}\right)
\pm n\pi\right].\lefteqn{}
\label{2.44}
\end{eqnarray}
It is interesting to note that when $T_1=T_2=T$ in the case of TDFC, 
the corresponding characteristic equation can be written as:
\begin{equation}
\lambda=A'+B'\,e^{-\lambda T},
\label{2.45}
\end{equation}
where $A'=A-K$ and $B'=B+K$. Noting the equivalency between Eq. (\ref{2.45}) 
and Eq. (\ref{2.3}), the exact analytical description of the stability 
domain in this case immediately follows from Theorem 1.

\section{Numerical example}

To test the VDFC method for stabilization of unstable steady 
states in chaotic RDDE systems, we will use the paradigmatic 
Mackey-Glass system introduced as a model for regeneration of blood cells 
in patients with leukemia \cite{MG77,FAR82,GP84,NPT95a,NPT95b}.
The Mackey-Glass equation in the presence of VDFC states:
\begin{equation}
\dot{x}(t)=\frac{a\,x(t-T_1)}{1+\left[x(t-T_1)\right]^c}-b\,x(t)+u(t),
\label{3.1}
\end{equation}
where $u(t)$ is given by Eqs. (\ref{2.4})--(\ref{2.5}). 
Here $x(t)$ is a concentration of circulating blood cells, and 
$a$, $b$ and $c$ are parameters of the 
free running system, involved in the description of 
the dependence of the 
production/destruction of the blood cells as a function 
of $x(t)$ and $x(t-T_1)$, respectively. We will consider the 
typical values $a=0.2$, $b=0.1$ and $c=10$. 

In the absence of control [$u(t)$=$0$], the system (\ref{3.1}) has 
a set of three fixed points $x_1^*=0$, $x_2^*=+1$ and $x_3^*=-1$ 
being solutions of:
\begin{equation}
\frac{a\,x^*}{1+{x^*}^c}-b\,x^*=0.
\label{3.2}
\end{equation}
The stability of each $x_i^*$ is obtained by linearizing the 
unperturbed system around $x_i^*$, leading to Eq. (\ref{2.2}) with:
\begin{equation}
A=-b,\hspace{1cm} B=a\,\frac{1+(1-c)\,{x^*}^c}{\left(1+{x^*}^c\right)^2},
\label{3.3}
\end{equation}
and the corresponding characteristic equation is given by Eq. (\ref{2.3}).
For $x_1^*=0$, we have $A=-b=-0.1$ and $B=a=0.2$. Using Theorem 1
we deduce that the fixed point $x_1^*$ is unstable for any $T_1$. 
For $x_{2,3}^*=\pm1$, we have $A=-b=-0.1$ and $B=a(2-c)/4=-0.4$, 
indicating that this pair of fixed points are
characterized by the same type of stability. From Theorem 1 we 
conclude that $x_{2,3}^*$ are stable if and only if $T_1\in[0,4.7082)$.
Figure 2 shows the trajectory of the unperturbed system in $x(t)$ vs. 
$x(t-T_1)$ coordinate space for four different values of $T_1$. 
\begin{figure}
\includegraphics[width=\columnwidth,height=!]{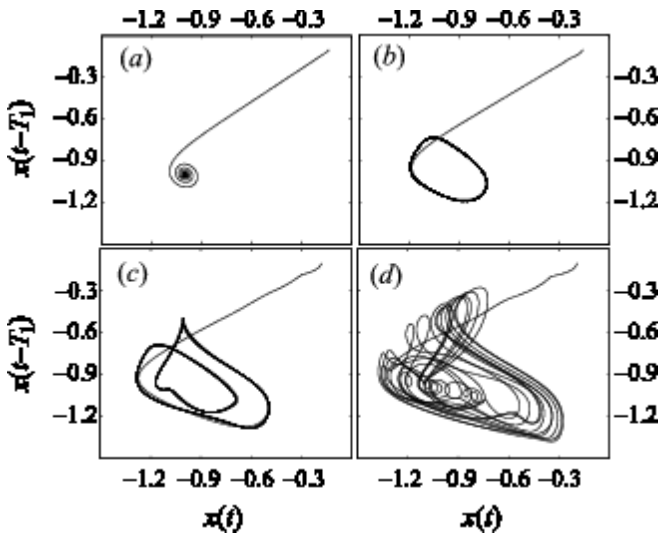}
\caption{Representative samples of phase plots 
$x(t)$ vs. $x(t-T_1)$ for the uncontrolled Mackey-Glass 
system at different values of $T_1$: (a) $T_1=4$ - the trajectory
is attracted to the stable equilibrium point $x_{3}^*=-1$; 
(b) $T_1=8$ - the trajectory approaches a limit cycle;
(c) $T_1=15$ - the attractor has evolved into a period-2 cycle; 
(d) $T_1=23$ - chaos. The simulations were performed using 
the MATLAB routine \texttt{dde23} for integrating delay-differential 
equations with constant delays.}
\end{figure}
Panel (a) shows the evolution of the system for $T_1=4$. 
Since for this value of $T_1$ 
the fixed points $x_{2,3}^*$ are stable, the preference 
of the system towards $x_{2}^*=+1$ or $x_{3}^*=-1$ depends 
on the initial conditions. Panels
(b)--(d) in Fig. 2 correspond to $T_1=8$, 15 and 23, respectively, 
showing the growth of the limit cycle through a period-doubling 
bifurcation sequence, and the eventual appearance of 
a chaotic attractor. 

In performing the stability analysis under VDFC, we will first 
consider a high-frequency modulation of the control delay $\tau(t)$.
The limitation imposed by Theorem 2 asserts that the fixed 
point $x_1^*$ cannot be stabilized with VDFC for any 
values of the control parameters $K$, $T_2$ and $\varepsilon$.
The validity of this assertion has been verified by the 
numerical simulations, showing the absence of the domains of 
successful control in the corresponding parametric planes.
On the other hand, the stability of the fixed points $x_{2,3}^*=\pm1$ 
is determined by the roots $\{\lambda_i\}$ of the 
characteristic Eq. (\ref{2.10}) with $A$ and $B$ given by 
Eq. (\ref{3.3}). 
Even though there exists an infinite number of roots 
$\lambda_i$ of Eq. (\ref{2.10}), only a finite number of them 
have real parts greater than a given constant.
A computation of the rightmost characteristic roots 
with large enough accuracy is a nontrivial nonlinear
eigenvalue problem, and there exist several effective methods 
to compute this part of the spectrum, e.g. by a discretization of either 
the time integration operator or the infinitesimal generator 
associated with the delay system \cite{BRE06,BMV04,BMV05,ENR02,VLR08}. 
Since the stability properties 
of the controlled system are determined by the characteristic roots 
with the leading real part, it is enough to employ a simple root-finding 
numerical algorithm based on the Newton-Raphson iteration method
with a suitable chosen grid of starting values.
For this purpose, we first make an implicit plot of the real and the 
imaginary parts of the characteristic Eq. (\ref{2.10}) in the 
complex $\lambda$ plane to visualize the approximate location of the roots 
as intersecting points between the corresponding curves. 
In this way we obtain a coarse estimate of the location of the rightmost
eigenvalues, the knowledge of which is then used to choose an 
appropriate grid of starting values encompassing this location.

By numerically solving Eq. (\ref{2.10}) with the aforementioned procedure, 
we obtain the domains of successful control in the parameter plane $(K,T_2)$ 
for a fixed delay $T_1$ and for different values of the 
amplitude $\varepsilon$. The results are shown in Figs. 3 and 4.
In the numerical calculations, we 
choose $T_1=23$ for which the original system is in a 
chaotic regime (see panel (d) in Fig. 2), having a positive 
value of the largest Lyapunov exponent (LLE = 0.00973) \cite{ljap}. 
\begin{figure}
\includegraphics[width=\columnwidth,height=!]{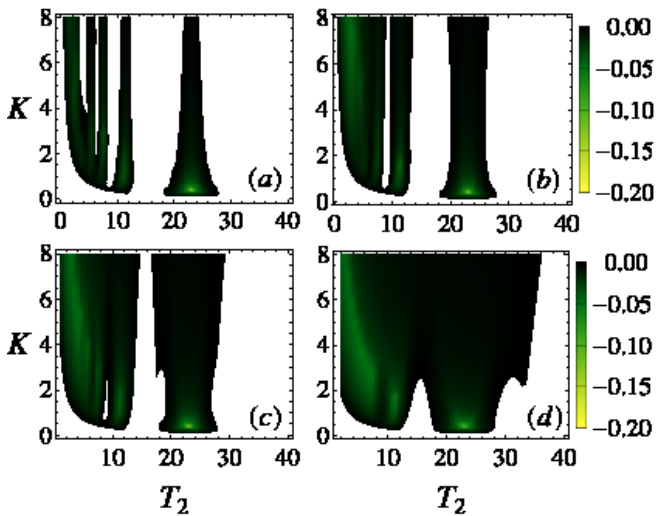}
\caption{(Color online) 
Domains of successful VDFC control in the $(K,T_2)$ plane for the
unstable equilibria $x_{2,3}^*=\pm1$ in the chaotic Mackey-Glass
system ($T_1=23$). The control delay $\tau(t)$ is modulated
with a sawtooth-wave, and the values of the modulation 
amplitudes are: (a) $\varepsilon=0$ (TDFC); 
(b) $\varepsilon=0.5$, (c) $\varepsilon=1$; (d) $\varepsilon=2$.
Combinations of $K$ and $T_2$ where VDFC successfully 
stabilizes the fixed points $x_{2,3}^*=\pm1$ are plotted in graytones
(colortones online). Note the shifts of the origin along the 
$T_2$ axes by an amount equal to $\varepsilon$ due to the limitation 
$T_2\geq\varepsilon$.}
\end{figure}
\begin{figure}
\includegraphics[width=\columnwidth,height=!]{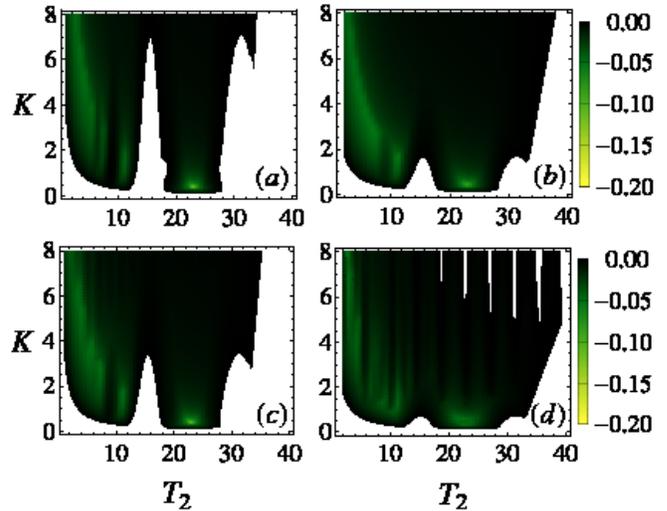}
\caption{(Color online) (a), (b) Stability domains of $x_{2,3}^*=\pm1$ 
in the ($K,T_2$) plane for the VDFC-controlled Mackey-Glass 
system with $T_1=23$. The delay modulation is in a form of 
a sine-wave with $\varepsilon=1$ (panel a) and $\varepsilon=2$ 
(panel b). (c), (d) Corresponding stability domains for a square-wave 
modulation. Note the shifts of the origin along the 
$T_2$ axes by an amount equal to $\varepsilon$ due to the limitation 
$T_2\geq\varepsilon$.}
\end{figure}
The shaded areas (color online) correspond to the set of control 
parameters $(K,T_2)$ for which the maximum of the real part 
of the characteristic eigenvalues $\{\lambda_i\}$ is negative
($\mathrm{max} [\mathrm{Re}\{\lambda_i\}]<0$), 
rendering the control successful. The values of 
$\mathrm{max} [\mathrm{Re}\{\lambda_i\}]$ are given by the 
grayscale (colorscale online) on the right in each figure, 
and the control is more robust as $\mathrm{max} [\mathrm{Re}\{\lambda_i\}]$ 
is more negative. The stability islands are 
surrounded by a "sea" of instability represented by the white 
region, for which the real part of the leading characteristic 
eigenvalue is positive ($\mathrm{max} [\mathrm{Re}\{\lambda_i\}]>0$). 
The "coastline" between stability and 
instability (the stability border) is given in a parametric 
form via Eqs. (\ref{2.29})--(\ref{2.30}) for $\varepsilon>0$ (VDFC),
and via Eqs. (\ref{2.43})--(\ref{2.44}) for $\varepsilon=0$ (TDFC).
Panels (a) through (d) of Fig. 3 correspond to the modulation  
of the feedback delay $\tau(t)$ in a form of a sawtooth-wave,
with amplitude values $\varepsilon=0$, 0.5, 1 and 2, respectively. 
Panel (a) reveals the structure of the stability domain 
for $\varepsilon=0$ (TDFC). For the
current choice of $T_1$, and also in general, there exists 
a stability region for relatively small $T_1$ with a complex
structure, and a resonance island encompassing $T_2=T_1=23$
for which the control is most robust and can be achieved 
with smaller values of $K$. As $\varepsilon$ becomes larger than
zero (VDFC, panels (b)--(d)), the structure of the stability domain
is reconfigured, resulting in a significant enlargement of the
area of successful control. This enlargement
is also observed for other delay modulations. In panels (a)--(b) of 
Fig. 4 we show the calculated stability domains for a sine-wave 
modulation for $\varepsilon=1$ and $2$, respectively, and 
(c)--(d) are the corresponding panels for a square-wave modulation. 
We note that for larger values of $\varepsilon$ in the case of a 
square-wave modulation, the stability area
eventually spreads into several clearly distinguished stability
islands, whose position is changing in an oscillatory manner
as $\varepsilon$ further increases.

In Fig. 5 we show the stability domains in $(T_1,T_2)$
plane, fixing the feedback gain value at $K=0.5$. 
\begin{figure}
\includegraphics[width=\columnwidth,height=!]{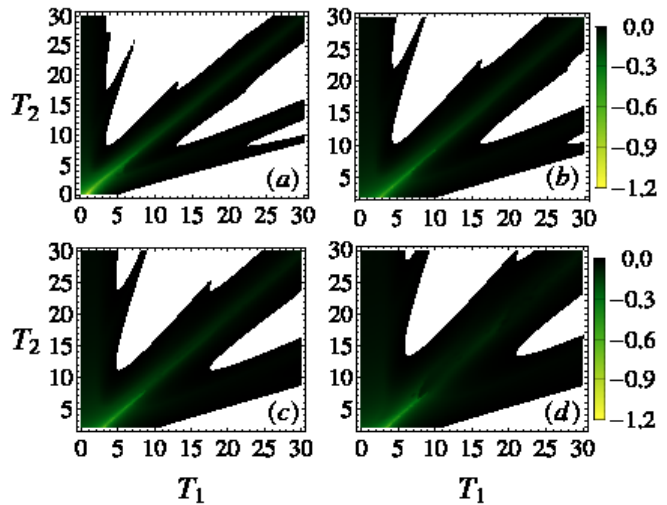}
\caption{(Color online) Domains of successful control 
in the ($T_1,T_2$) plane for the unstable fixed points $x_{2,3}^*=\pm1$
in the Mackey-Glass system. The feedback gain is fixed at $K=0.5$. 
(a) Stability diagram for $\varepsilon=0$ (TDFC). 
(b)--(d) Respective stability diagrams for sawtooth, 
sine and square-wave modulations with
$\varepsilon=2$ (VDFC). Note that the minimum value of the 
$T_2$-axis in panels (b)--(d) is $T_2=2$ due to the limitation 
$T_2\geq\varepsilon$.}
\end{figure}
Panel (a) depicts the case when the modulation 
is absent (TDFC, $\varepsilon=0$),
and panels (b)--(d) are related to sawtooth, sine and square-wave 
modulations, respectively, with $\varepsilon=2$. The diagrams show
the typical enlargement of the stability area for VDFC with respect 
to TDFC. The parametric representation of the stability boundary is 
given by Eqs. (\ref{2.34})--(\ref{2.35}).

To verify the analysis in the previous paragraphs, we performed a
computer simulation of VDFC for the fixed points $x_{2,3}^*=\pm1$ 
by numerically integrating the system (\ref{3.1}) 
for different delay modulations.
The results are shown in Fig. 6. 
\begin{figure*}
\includegraphics[width=\textwidth,height=!]{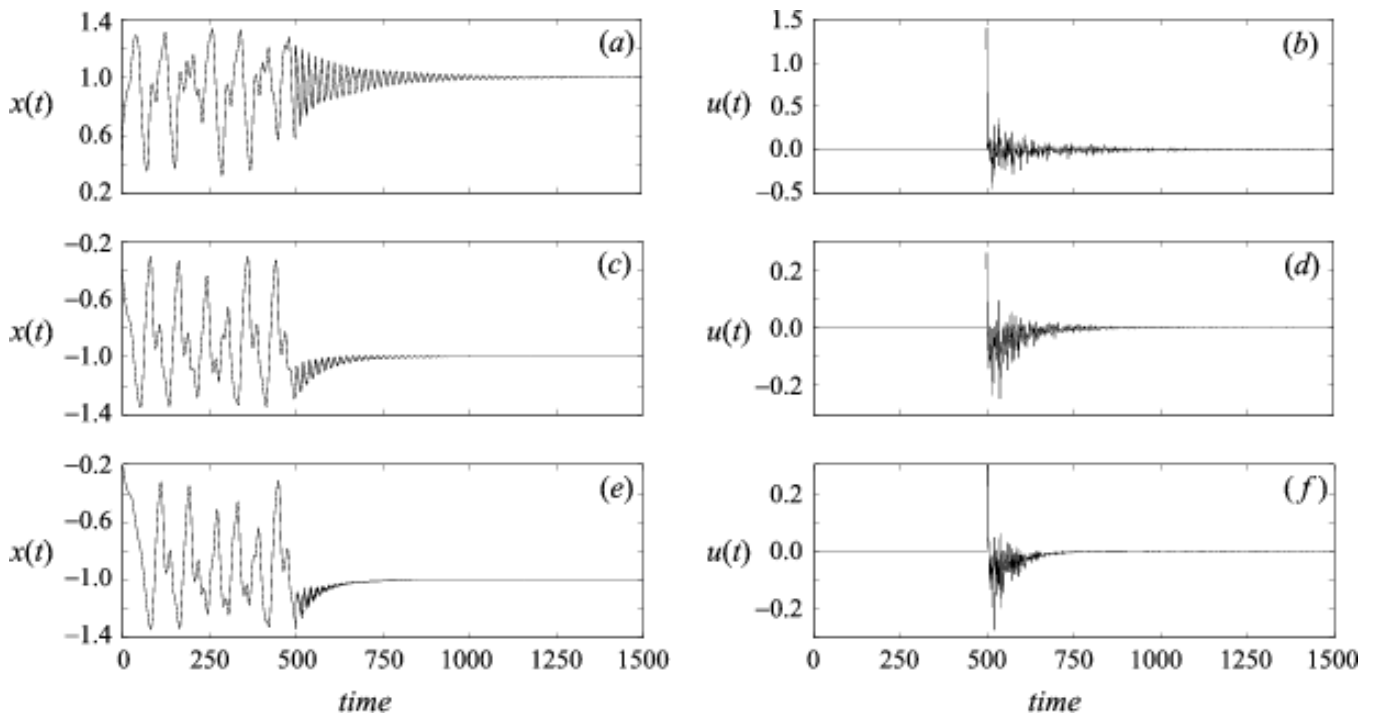}
\caption{VDFC applied to the chaotic Mackey-Glass system 
using different modulations of the delay-time $\tau(t)$. 
The parameters of the uncontrolled system are: 
$a=0.2$, $b=0.1$, $c=10$, $T_1=23$.
(a), (b) Time plots of the variable $x(t)$ and the feedback 
signal $u(t)$ for a sawtooth-wave modulation, indicating a successful
control of the unstable fixed point at $x_2^*=+1$. (c), (d)
Stabilization of the unstable equilibrium at $x_3^*=-1$
with a sine-wave modulation. (e), (f) Time-series for
a square-wave modulation stabilizing the unstable point at
$x_3^*=-1$. In each case, the control parameters were: 
$K=2$, $\varepsilon=2$, $T_2=18$ and $\nu=5$. The control
was activated at $t=500$. The total time span shown in each
panel is 1500 time units. The simulations were performed using
the MATLAB routine \texttt{ddesd} for integrating delay-differential 
equations with general delays.}
\end{figure*}
Panels (a), (c) and (e) 
depict the dynamics of the variable $x(t)$ for sawtooth, 
sine and square-wave modulations, respectively, and 
panels (b), (d) and (f) show the corresponding time-series of the
feedback signal $u(t)$. In each case, the control parameters 
were chosen as $K=2$, $\varepsilon=2$, $T_2=18$ and $\nu=5$,
fixing the delay of the uncontrolled system at $T_1=23$
for which the system is chaotic. 
We note that for these parameter values, the control 
via TDFC ($\varepsilon=0$) is unsuccessful for any $K$, 
as can be perceived from the stability domain depicted in
panel (a) of Fig. 3. 
Also, since $x_{2,3}^*$ have identical set of 
characteristic eigenvalues, they share common domains 
of successful control. However, they have different basins 
of attraction, and the preference of control towards 
either $x_2^*$ or $x_3^*$ depends on the initial conditions.
In panels (b), (d) and (f) we see that the feedback signal
$u(t)$ vanishes when the stabilization of the fixed point
is achieved, suggesting noninvasiveness of VDFC, which is a
consequence of the form of the control force in Eq. 
(\ref{2.4}), since $x(t-\tau(t)) = x(t)$ if the fixed point 
is stabilized. 

When the frequency $\nu$ of the delay modulation
is below the threshold (low-frequency modulation), 
the approximation of the variable-delay system with a 
distributed-delay system is not covered by Theorem A1,
and, hence, the control domains cannot be calculated from
the characteristic Eq. (\ref{2.10}).
However, the stability domains in this case
can be obtained by numerically integrating the linear 
variable-delay system (\ref{2.7})--(\ref{2.8}) for 
different values of the corresponding control parameters. 
In Fig. 7 we show the results of such a simulation in 
the parametric plane $(K,T_2)$  for $T_1=23$ and 
$\varepsilon=2$, taking the time-modulation of $\tau(t)$ 
in a form of a sawtooth wave. Different panels of the figure 
correspond to different values of the delay 
frequency $\nu$: (a) $\nu=1.4$, (b) $\nu=1.8$, (c) $\nu=1.9$,
(d) $\nu=2.0$, (e) $\nu=2.2$, (f) $\nu=2.4$, (g) $\nu=2.6$, 
(h) $\nu=3.0$. The combinations $(K,T_2)$ leading to
a successful stabilization of the unstable fixed points
$x_{2,3}^*=\pm1$ are marked in black.
It is observed that when the modulation frequency is 
about $\nu=3.0$ (panel (h)), the structure of the 
stability domain fairly resembles the stability domain
for a high-$\nu$ modulations obtained from the characteristic 
Eq. (\ref{2.10}) (compare with panel (d) in Fig. 3, noting the 
different scales on the $T_2$ axis). As expected, the 
simulations show that this resemblance becomes improved 
as $\nu$ attains higher values. On the other hand, the 
structure of the stability domain is gradually changing 
as $\nu$ becomes smaller than $\nu=3.0$ (panels (a)--(g)), 
resulting in a reconstruction of the main domain and a
birth of many small stability islands, clearly notable 
for larger nominal delays $T_2$ and approximately 
centered about those $T_2$ which are odd multiples of $\pi/\nu$.
The emergence of this additional domain structure 
could be due to a resonance between the delay
frequency $\nu$ and the intrinsic frequencies of the 
uncontrolled system, which are infinite in number.
The distance between these resonance islands 
($\approx 2\pi/\nu$) becomes wider as $\nu$ decreases, 
and they become less pronounced for lower values of $\nu$.
It can be noticed that the appearance of these 
resonance islands allows stabilization of the unstable 
equilibria for much larger nominal delays $T_2$ 
in comparison to the values of $T_2$ for a high-$\nu$ 
modulation. The simulations show that the range 
of the delay frequency parameter containing these 
resonance islands is strongly dependent on the 
system parameters (e.g., the modulation amplitude) 
and on the type of the delay modulation, and that this 
range of $\nu$ may not be continuous as in the current 
case, but it may consist of several different subintervals 
spread throughout the entire $\nu$-interval below some
sufficiently high frequency and encompassing some of the values of $\nu$ 
coinciding with the eigenfrequencies of the uncontrolled 
system (see Fig. 8). 

To check if the limitation of the control method asserted by 
Theorem 2 remains valid for low-frequency 
modulations, we have performed numerical simulations to 
determine the domains of successful VDFC control of the 
unstable equilibrium $x_1^*=0$, which has 
been shown uncontrollable via Theorem 2 for high-frequency 
modulations. The simulations in this case show the 
absence of the control domains in the corresponding 
parametric planes, suggesting the validity of Theorem 2 
in the entire frequency range. 

\begin{figure*}
\includegraphics[width=\textwidth,height=!]{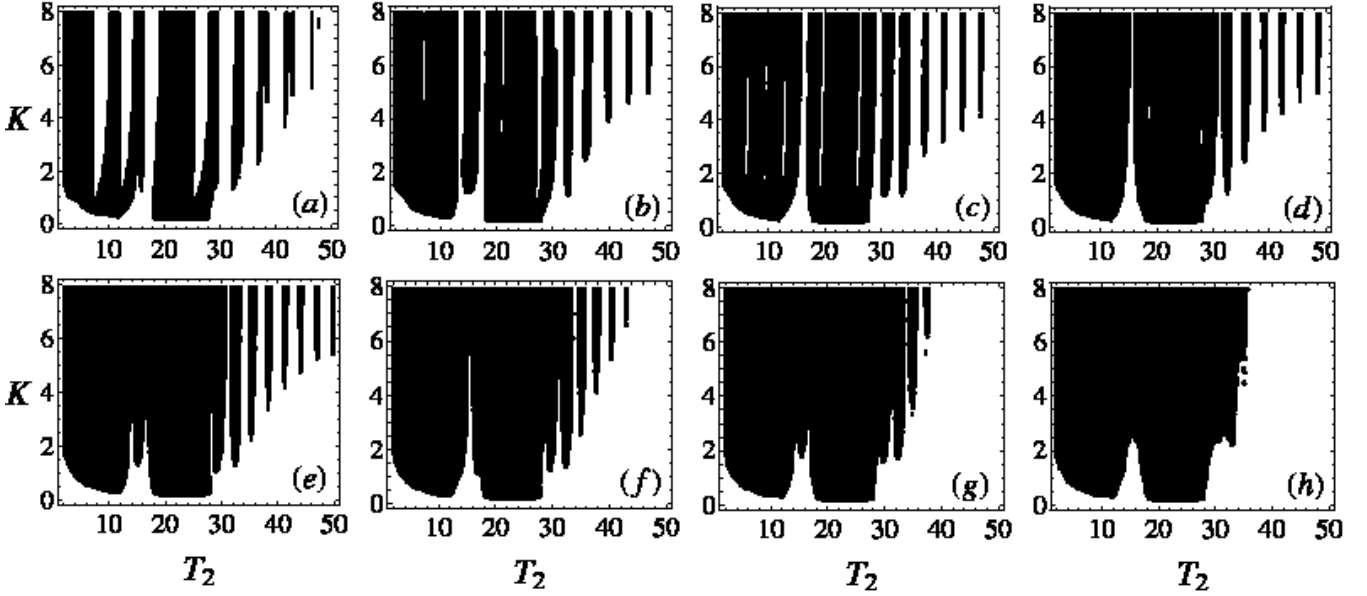}
\caption{
The stability domains in $(K,T_2)$ plane related 
to the unstable steady states $x_{2,3}^*=\pm1$ of the 
chaotic Mackey-Glass system $(T_1=23)$ for a low frequency modulation of the 
time delay $\tau(t)$. The delay modulation is in a form 
of a sawtooth-wave with $\varepsilon=2$, and the value 
of the modulation frequency  is: (a) $\nu=1.4$, (b) $\nu=1.8$, 
(c) $\nu=1.9$, (d) $\nu=2.0$, (e) $\nu=2.2$, (f) $\nu=2.4$, 
(g) $\nu=2.6$, (h) $\nu=3.0$. The combinations $(K,T_2)$ 
leading to a successful stabilization of the unstable 
equilibria $x_{2,3}^*=\pm1$ are marked in black.
The characteristic eigenfrequencies of the uncontrolled
system that lie in this interval of $\nu$ are:
1.43, 1.71, 1.98, 2.25, 2.52 and 2.80. Note the appearance 
of the resonance islands at the right of the main structure
as $\nu$ becomes smaller than $\nu=3.0$. Also note the 
shifts of the origin along the $T_2$ axes by an amount equal to 
$\varepsilon$ due to the limitation $T_2\geq\varepsilon$.
}
\end{figure*}
\begin{figure*}
\includegraphics[width=\textwidth,height=!]{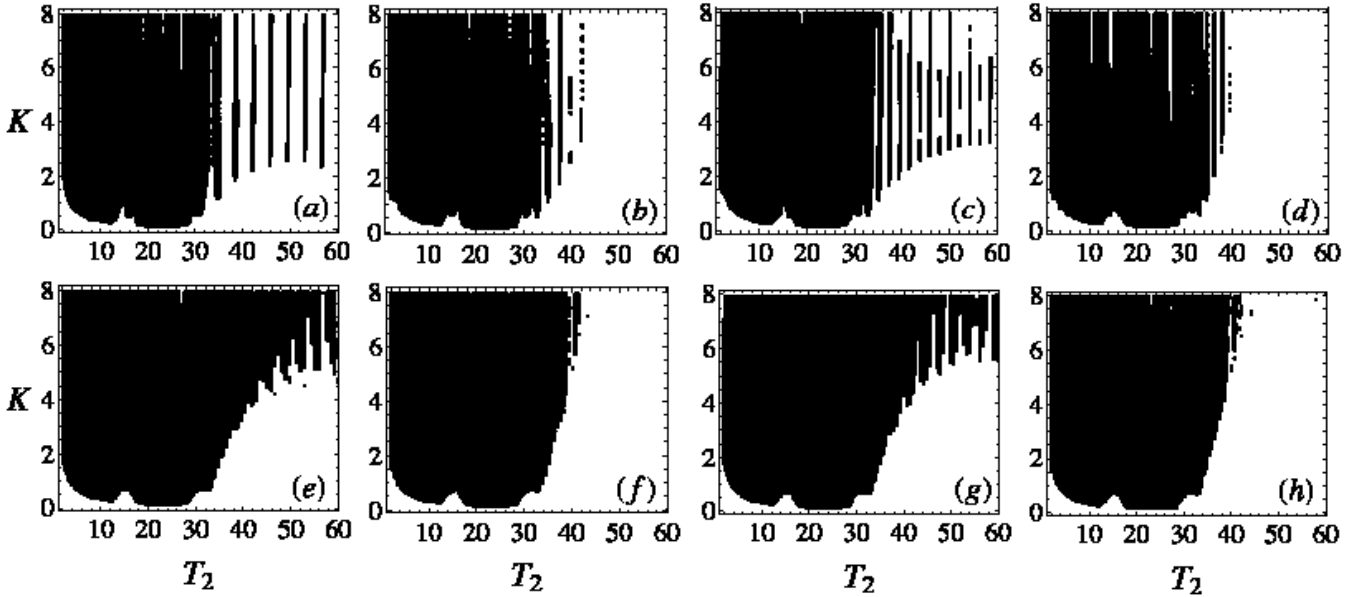}
\caption{
A sample of the low-frequency control domains in $(K,T_2)$ plane 
corresponding to the unstable steady states $x_{2,3}^*=\pm1$ of the 
chaotic Mackey-Glass system $(T_1=23)$ for a square-wave modulation 
with $\varepsilon=2$. The value of the modulation frequency 
is: (a) $\nu=1.7$, (b) $\nu=2.8$, 
(c) $\nu=3.0$, (d) $\nu=4.0$, (e) $\nu=8.0$, (f) $\nu=10.0$, 
(g) $\nu=11.0$, (h) $\nu=12.0$. 
The range of $\nu$ in which the resonance islands exist 
consists of several distinguished intervals encompassing
some of the eigenfrequencies of the uncontrolled system:
1.71, 3.07, 7.99, 10.99. Note that although the modulation
frequency $\nu$ in panel (b) coincides with one of the 
eigenfrequencies of the uncontrolled system ($\approx2.8$), the 
resonance islands are hardly noticeable in this case.
}
\end{figure*}

\section{Conclusion}

In conclusion, we have shown that variable-delay feedback 
control allows stabilization of unstable 
steady states in a class of scalar retarded time-delayed systems,
represented by the chaotic Mackey-Glass system, 
over much larger domain of parameters in comparison to the 
usual Pyragas' delayed feedback control scheme. 
The analysis showed that the enlargement
of the control domain may undergo a complex rearangement 
depending on the type and the frequency of the delay modulation. 
It is noticed that the enlargement of the control domain 
for high-frequency modulation of the delay is more 
pronounced when the variable delay is a continuous function 
of time in contrast to the case of variable delay function 
with a discontinuity leading to complex stability 
domain structure of a lesser magnitude. In the case of
low-frequency modulation of the delay, we notice 
a complex rearrangement of the control domain,
resulting in an appearance of extra stability islands, probably
a consequence of a resonance between the frequency of the variable 
delay and the eigenfrequencies of the uncontrolled system.
This resonance effect allows successful stabilization 
of the unstable fixed point for much larger nominal 
delays with respect to the situation when the frequency 
of the delay variation is above the threshold.

Limitation imposed by Theorem 2 shows that VDFC method 
fails to control certain unstable steady states for any value 
of the feedback control parameters in the case when the frequency
of the delay modulation is high. 
Moreover, numerical simulations suggest that this 
limitation is also valid for low-frequency modulations.
Nevertheless, in lack of any analytical tool to treat a 
low-frequency modulated VDFC, any general statement 
concerning the generalization of Theorem 2 to the whole 
frequency range should be taken cautiously, as well as
the related observations concerning the aforementioned
resonance phenomenon. 

Putting the observations related to low-frequency modulation 
of the control delay on a firm mathematical basis 
constitutes an interesting subject for a future study. 
Other possible directions for future consideration would be 
stabilization of unstable steady states by VDFC in 
other types of DDE systems (e.g. systems described by 
neutral delay-differential equations \cite{BKH08}),
in systems described by partial differential equations,
and, also, implementation of the control method to stabilize 
unstable periodic orbits by a suitable choice of the delay 
modulation in order for the control method to stay 
noninvasive. An example for such a modulation in the
latter case would be a periodic change of the 
control delay between $T$ and $2T$, where $T$ is the
period of the orbit to be stabilized \cite{GU09}.

\acknowledgments

We thank one of the referees for valuable comments that 
improve the presentation in the paper, in particular
the discussion related to the resonance phenomenon
in low-frequency modulations. We also appreciate
fruitful discussions with B. Fiedler on the limitations of
the "odd-number limitation".

\appendix
\section{}

The stability of a linear (or linearized) RDDE system with a 
fast-varying delay can be obtained by studying the 
roots of the characteristic equation of the related time-invariant 
distributed delay system. The correctness of this 
approach is guaranteed only if the frequency of variation 
of the delay is large compared to the system's dynamics. 
A precise formulation of these assertions constitutes the following 
theorem:

\newtheorem{theorema}{Theorem}[section]
\begin{theorema}
Consider the linear system of variable delay differential equations:
\begin{eqnarray}
\frac{d}{dt}\mathbf{x}(t)&=&\mathbf{\hat{A}}\cdot\mathbf{x}(t)+\mathbf{\hat{B}}\cdot\mathbf{x}(t-T_1)+\mathbf{u}(t),
\label{A1}\\
\mathbf{u}(t)&=&\mathbf{\hat{K}}\cdot(\mathbf{x}(t-T(t))-\mathbf{x}(t)),
\label{A2}\\
T(t)&=&T_2+\varepsilon f(\nu t),
\label{A3}
\end{eqnarray}
where $\mathbf{\hat{A}},\mathbf{\hat{B}},\mathbf{\hat{K}}\in\mathbb{R}^{N\times N}$ are constant matrices, 
$\mathbf{x}(t)\in\mathbb{R}^{1\times N}$, and $f:\mathbb{R}\rightarrow[-1,1]$
is a periodic function with zero mean and period $2\pi$, $\mathrm{max} f=1$,
and $\mathrm{min} f=-1$. Let $\varepsilon,T_2,\nu\in\mathbb{R}_0^+$, and
$\varepsilon\leq T_2$. Let the integrable function $w:[-1,1]\rightarrow
\mathbb{R}^+$ be defined by:
\begin{equation}
\int_{-1}^{1}\alpha(t)w(t)dt=\frac{1}{2\pi}\int_0^{2\pi}\alpha(f(t))dt
\label{A4}
\end{equation}
for every continuous function $\alpha:[-1,1]\rightarrow\mathbb{R}$. If the 
comparison system:
\begin{eqnarray}
\lefteqn{\frac{d}{dt}\mathbf{x}(t)=\mathbf{\hat{A}}\cdot\mathbf{x}(t)+\mathbf{\hat{B}}\cdot\mathbf{x}(t-T_1)+}\nonumber\\
&+&\mathbf{\hat{K}}\cdot\left(
\displaystyle{\int_{t-T_2-\varepsilon}^{t-T_2+\varepsilon} 
\frac{w((\theta-t+T_2)/\varepsilon)}{\varepsilon}\mathbf{x}(\theta)\,d\theta}
-\mathbf{x}(t)\right)\nonumber\\
\label{A5}
\end{eqnarray}
is asymptotically stable, then the original system (\ref{A1})--(\ref{A3})
is globally uniformly asymptotically stable for large values of the 
frequency $\nu$ of the modulation.
\end{theorema}

Theorem A1 is a restatement of the main result in Ref. \cite{MAN05} 
to accomodate the present discussion \cite{MIC}, and its proof is based on an 
extension of the recently introduced trajectory-based proof 
technique \cite{MA00}. 
According to Theorem A1, the stability of (\ref{A1}) under the 
variable-delay control force (\ref{A2}) can be inferred from the 
stablilty of the analogous time-invariant system 
(\ref{A5}) with a distributed delay, for sufficiently large values
of the parameter $\nu$ determining the frequency of the modulation. 
It is worth noting that Theorem A1 can be generalized to include 
the most general case of multiple delayed feedback terms in 
the control force (\ref{A2}) with different types of delay 
modulations \cite{MIC}. 
The proof of this extension is straightforward, 
following the lines of the proof given in \cite{MAN05}.

The comparison system (\ref{A5}) can be recast in the form:
\begin{widetext}
\begin{equation}
\frac{d}{dt}\mathbf{x}(t)=\mathbf{\hat{A}}\cdot\mathbf{x}(t)+\mathbf{\hat{B}}\cdot\mathbf{x}(t-T_1)
+\mathbf{\hat{K}}\cdot\left(
\int_{-1}^{1}w(\eta)\mathbf{x}(\varepsilon\eta+t-T_2)\,d\eta
-\mathbf{x}(t)\right),
\label{A6}
\end{equation}
\end{widetext}
by making a change of the integration variable $\theta$ to the new 
variable $\eta$ through the relation $\theta=\varepsilon\eta+t-T_2$.
Furthermore, by taking $\alpha(t)=1$ and $\alpha(t)=t$ in Eq. (\ref{A4})
respectively, we obtain the relations involving the weight function $w$:
\begin{eqnarray}
\int_{-1}^{1}w(t)dt=1, \label{A7}\\
\int_{-1}^{1}tw(t)dt=0. \label{A8}
\end{eqnarray}
From Eq. (\ref{A4}), the weight $w$ can be interpreted as the
probability distribution of $f(\xi)$, where $\xi$ is uniformly
distributed over the interval $[0,2\pi]$ (see Table I).

\end{document}